\documentclass[a4paper]{article}

\usepackage{INTERSPEECH2019}
\usepackage{amsmath,graphicx,indentfirst,url,dirtree,color}
\usepackage{ascmac}
\usepackage{CJKutf8}
\usepackage{multirow}

\newcommand{\jp}[1]{ \begin{CJK}{UTF8}{ipxm}#1\end{CJK} }

\newlength\savedwidth
\newcommand{\wcline}[1]{\noalign{\global\savedwidth\arrayrulewidth\global\arrayrulewidth 1.0pt} \cline{#1}
\noalign{\global\arrayrulewidth\savedwidth}}

\title{Construction and Analysis of Impression Caption Dataset\\ for Environmental Sounds}

\name{Yuki Okamoto$^{1*}$, Ryotaro Nagase$^{2*}$, Minami Okamoto$^2$, Yuki Saito$^1$,\\ Keisuke Imoto$^3$, Takahiro Fukumori$^2$, and Yoichi Yamashita$^2$}
\address{$^1$The University of Tokyo, Japan \\
$^2$Ritsumeikan University, Japan \\$^3$Doshisha University, Japan}

\email{y-okamoto@ieee.org}

\begin{document}
\maketitle
\begin{abstract}
Some datasets with the described content and order of occurrence of sounds have been released for conversion between environmental sound and text.
However, there are very few texts that include information on the impressions humans feel, such as ``sharp'' and ``gorgeous,'' when they hear environmental sounds.
In this study, we constructed a dataset with impression captions for environmental sounds that describe the impressions humans have when hearing these sounds.
We used ChatGPT to generate impression captions and selected the most appropriate captions for sound by humans.
Our dataset consists of 3,600 impression captions for environmental sounds.
To evaluate the appropriateness of impression captions for environmental sounds, we conducted subjective and objective evaluations.
From our evaluation results, we indicate that appropriate impression captions for environmental sounds can be generated. 
\end{abstract} \vspace{1mm}


\section{Introduction}
\renewcommand{\thefootnote}{\fnsymbol{footnote}}
\footnotetext[1]{These authors contributed equally to this work.}
\renewcommand{\thefootnote}{\arabic{footnote}}
Research in environmental sound analysis and synthesis using deep learning has been actively pursued \cite{Choi_DCASE2023_01,mesaros_IEEE2021_01}.
With the development of a large language model (LLM), tasks that enable interconversion between environmental sounds and text, such as describing the content of environmental sounds in natural language (audio captioning) \cite{drossos_WASPAA2017_01,kim_ICASSP2023_01} and artificially generating environmental sounds from natural language (text-to-audio) \cite{liu_TASLP2024_01,Kreuk_ICLR2023_01,okamoto_ATSIP_2022}, have gained attention.
The mutual conversion technology between environmental sounds and text has potential applications in various fields, such as media content production.

A large number of sound--text pairs are required for mutual conversion between environmental sounds and text by a statistical approach.
Thus, several datasets of environmental sound and text pairs have been released \cite{kim_audiocaps,Drossos_ICASSP2020_01}.
For example, AudioCaps \cite{kim_audiocaps}, created for audio captioning, contains approximately 50,000 environmental sound--text pairs.
Another dataset built using LLM, WavCaps \cite{mei_TASLP2024_01}, contains approximately 400,000 environmental sound--text pair data.
However, the descriptions of the environmental sounds in these datasets are limited to the content and order of occurrence of the sounds, e.g., ``men talking, different birds singing at the same time.''
In particular, very few texts include impression information, such as ``sharp'' and ``gorgeous'' that humans feel when they hear environmental sounds.
If the impression information of environmental sounds can be utilized, it can lead to a technology for recommending and automatically generating environmental sounds in accordance with the impressions given to content consumers when creating media content.
Moreover, the use of impression information for environmental sounds can be expected to lead to a more expressive understanding of audio captioning.

In this study, we constructed a dataset with impression captions for environmental sounds that describe the impressions humans have when hearing these sounds.
First, we collected impression words for environmental sounds via a crowdsourcing service.
Second, we generated impression captions in Japanese for environmental sounds by a large language model using collected impression words generated.
Finally, we selected the most appropriate impression caption for an environmental sound through a crowdsourcing service.

The rest of the paper is organized as follows.
In Sec.~\ref{creation_dataset}, we describe the creation of the dataset.
In Sec.~\ref{analysis_dataet}, we discuss the analysis of our dataset.
Finally, we summarize and conclude this paper in Sec.~\ref{conclusion}.
 
\section{Creation of dataset}
\label{creation_dataset}
We constructed a dataset of impression captions for environmental sounds in two stages: the collection of impression words for environmental sounds (Sec.~\ref{collect_impression_words}) and the generation of impression captions by ChatGPT and the selection of appropriate captions for sounds by humans (Sec.~\ref{generate_caption}).
It is also possible to collect impression captions directly from humans.
However, if impression captions are collected directly from a human, each impression caption may include more than just a description of the impression, such as the sound event label.
Thus, we used ChatGPT to generate impression captions.
In this study, we collected impression words and generated impression captions for the environmental sounds of ESC-50 \cite{Piczak_CoM2_01}.
ESC-50 has five major categories, each having 10 sound events and 40 sounds for each sound event. 
In this paper, we used 1,200 sounds in three categories: natural soundscapes, water sounds, interior/domestic sounds, and exterior/urban noises.

\subsection{Design of our dataset}
Our dataset consists of the following contents:

\begin{itemize}
  \item Impression words for environmental sounds\\
  We collected a total of 3,600 impression words (three impression words $\times$ 1,200 environmental sounds) from crowdworkers for environmental sounds in ESC-50.\\
  \item Confidence score for each impression word\\
  We collected a total of 3,600 confidence scores from crowdwokers, who themselves transcribed the caption.\\
  \item Impression captions for environmental sounds\\
  We generated impression captions for environmental sounds using ChatGPT API.
  We also collected 3,600 impression captions (three impression captions $\times$ 1,200 environmental sounds) most appropriate for the environmental sounds by humans through a crowdsourcing service from the candidates generated by ChatGPT API.\\
  \item Appropriateness score for each impression caption\\
  We collected appropriateness scores for each impression caption from three crowdworkers who did not transcribe impression captions.\\
\end{itemize}

\begin{figure}[t!]
\centering
\includegraphics[scale=1.1]{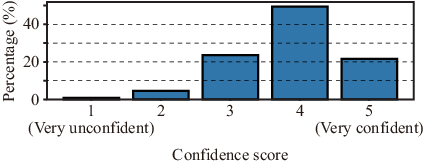}
\caption{Histogram of confidence score}
\label{fig:confidence_hist}
\end{figure}
\subsection{Collection of impression words}
\label{collect_impression_words}
Using a crowdsourcing service, we collected impression words in Japanese for environmental sounds.
Crowdsourcing services can collect sound impression words from a large number of workers, allowing us to gather a wide variety of impression words for each sound.
We collected a total of 1,200 environmental sounds from 30 sound events included in ESC-50.
We believe that it is difficult to assign impression words to sounds made by living things, such as animal sounds and people sneezing.
Therefore, we collected impression words for nonliving environmental sounds.
Impression words were collected for a total of 1,200 environmental sounds (30 sound events $\times$ 40 sounds) included in ESC-50.
We collected impression words from three workers per environmental sound. 
We presented only the sounds to crowdworkers to eliminate the bias derived from sound event labels and other information, and asked crowdworkers to express their impressions in a free-text format.
To collect only impression words for the environmental sound, we instructed the workers to not describe sound events or use onomatopoeic words.
We also collected a 5-scale confidence score from 1 (very unconfident) to 5 (very confident) for the impression words for environmental sounds from workers.

In the collected impression words, some captions included the names of sound events.
To remove these, we conducted a morphological analysis using MeCab, which is an open-source text segmentation library for Japanese written text, and excluded captions containing words inappropriate for expressing impressions, such as nouns.
We extracted only those captions with adjectives, adjectival verbs, verbs, and adjectival verb stems considered appropriate for expressing impressions of environmental sounds from morphological analysis results.

Figure \ref{fig:confidence_hist} shows the histogram of confidence score for the collected impression words. 
The figure shows that the overall confidence level tended to be high.
This result indicates that assigning impression words to environmental sounds was relatively easy.
\begin{figure}[t!]
\centering
\includegraphics[scale=0.95]{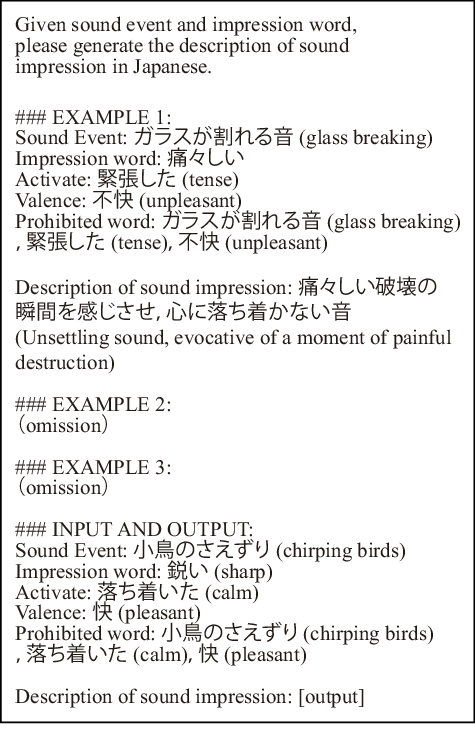}
\caption{Examples of giving instructions to ChatGPT}
\label{fig:promt_example}
\end{figure}
\begin{table*}[t!]
\caption{Examples of impression captions}
\label{table:example_caption}
\centering
\resizebox{\linewidth}{!}{%
\begin{tabular}{@{}lccc|l@{}}
    \wcline{1-5}
    \\[-8pt]
    \multicolumn{4}{c|}{Input to ChatGPT} & \multirow{3}{*}{Output impression caption}\\
    \cline{1-4}
    \\[-8pt]
     \multirow{2}{*}{Sound event} & Pleasant or & Calm or  & Impression & \\ 
      &  unpleasant & tense & word & \\ 
    \wcline{1-5}
    \\[-8pt]
      Chirping birds & Pleasant & Calm &  Sharp & Impression of sharp and clear sound reaching the ear and bringing calm and peace\\
     \jp{（小鳥のさえずり）}& \jp{（快）} & \jp{（落ち着いた）} & \jp{（鋭い）} & to the mind \jp{（鋭く澄んだ音が耳に届き、心に穏やかな安らぎをもたらす印象）}\\
     \cline{1-5}
      Church bells & Pleasant & Tense & Gorgeous & Gorgeous tones ring out, joyful sounds that foretell of new beginnings\\
     \jp{（教会の鐘の音）}& \jp{（快）} & \jp{（緊張した）} & \jp{（華やかな）} & \jp{（華やかな音色が鳴り響き、新たな始まりを予感させる喜びの音）}\\
     \cline{1-5}
     Door knock & Unplesant & Calm &  Scary & Its heavy sound of fear brings tension and anxiety to the mind\\
    \jp{（ドアを叩く音）} & \jp{（不快）} & \jp{（落ち着いた）} & \jp{（怖い）} & \jp{（怖さを感じさせるその重い響きは、心に緊張と不安をもたらす）}\\
    \cline{1-5}
    Vacuum cleaner & Unpleasant & Tense & Noisy & Noise that echoes loudly and instantly disturbs the mind\\
     \jp{（掃除機の音）}& \jp{（不快）} & \jp{（緊張した）} & \jp{（うるさい）} & \jp{（うるさく響き、一瞬で心をざわつかせる雑音）}\\
     \wcline{1-5}
\end{tabular}
}
\end{table*}

\subsection{Generation of impression captions by ChatGPT and selection of appropriate captions by humans}
\label{generate_caption}
Referring to the paper by Nagase et al. \cite{Nagase_ASJ_01}, we generated impression captions using collected impression words in Sec.~\ref{collect_impression_words} using ChatGPT.
We used the OpenAI ChatGPT API (``gpt-4o'') to generate impression captions in two steps.
The data generation period was May 2024.

{\bf Step 1 Creation of impression caption candidates:} 
We generated candidate impression captions by ChatGPT.
LLM has been utilized in dataset construction and expansion \cite{fang_arXiv2023_01}, and in WavCaps, a dataset of environmental sound-text pairs, LLM is used to paraphrase expressions in explanatory sentences. 
There are also studies in which LLM is utilized to efficiently generate text containing information on human emotions \cite{Xin_IEEE_2024}.
Therefore, we believe that LLM is also effective for generating impression captions in this study.

Figure \ref{fig:promt_example} shows examples of giving instructions to ChatGPT.
We provided only Japanese sentences for sound event labels provided by ESC-50, impression words collected in Sec.~\ref{collect_impression_words}, and emotional impressions for input to ChatGPT.
The instructions included ``Please write down your impression of the sound based on the given sound events and impression words'', sample responses, sound event labels, and impression words given to ChatGPT, and whether they were the emotional impression of ``pleasant-calm'', ``pleasant-tense'', ``unpleasant-calm'', or ``unpleasant-tense''. 
For example, impression words such as ``sharp'' may have emotional impression, such as a positive or negative impression. 
Thus, in addition to impression words, we used emotional impressions such as "pleasant-calm" for input into ChatGPT to generate diverse impression captions.
The generated impression caption did not include the emotional impression used for input or the names of sound events.
We generated 100 impression captions for each emotional impression.

{\bf Step 2 Selection of the most appropriate impression caption for the environmental sound from the candidates:} We selected the most appropriate impression caption from the candidates created in Step 1 by humans through crowdsourcing.
We presented four sentences from impression caption candidates to crowdworkers, one randomly for each of the classes "pleasant-calm," "pleasant-tense," "unpleasant-calm," and "unpleasant-tense. 
The crowdworkers selected what they considered was the most appropriate impression caption for the environmental sound from the four sentences presented.
Five workers selected an impression caption for each sound, and the impression caption for the environmental sound was decided by a majority vote.
If a majority vote did not result in a decision, a random selection was made from the two options that received the most responses, and this was used as the impression caption for the environmental sound. 
As a result of Step 2, we collected a total of 3,600 sentences (three captions per environmental sound).
Table \ref{table:example_caption} shows examples of the impression captions.

\begin{figure}[t!]
\centering
\includegraphics[scale=1.0]{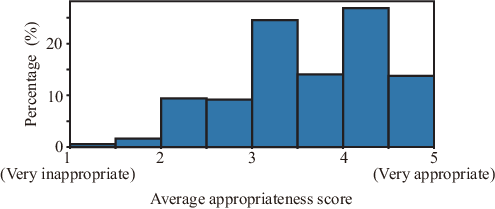}
\caption{Histogram of average appropriateness score for each impression caption}
\label{fig:appropriateness_hist}
\end{figure}

\section{Analysis of our dataset}	
\label{analysis_dataet}
As an analysis of the constructed dataset, we conducted a subjective evaluation of impression captions for environmental sounds as described in Sec.~\ref{subject_eval}, and evaluations of text-to-audio retrieval and audio-to-text retrieval as described in Sec.~\ref{object_eval}.
If the retrieval model can be trained using the data from the impression caption and environmental sound pairs, we can consider that the impression caption is valid for the environmental sound.

\subsection{Subjective evaluation}
\label{subject_eval}
We conducted a subjective evaluation to determine the appropriateness of the impression captions generated by ChatGPT for environmental sounds.
We used a crowdsourcing service for the subjective evaluation.
Each pair of environmental sound and impression caption was evaluated by three crowdworkers.
The crowdworkers were presented with the environmental sound and the impression caption.
They scored the appropriateness of the impression caption for the presented environmental sound on a 5-point scale from 1 (very inappropriate) to 5 (very appropriate).
We evaluated all collected environmental sound-impression caption pairs.

Figure \ref{fig:appropriateness_hist} shows the results of the average appropriateness of the impression captions for the environmental sounds.
The scores in the figure are the average of the appropriateness scores of each environmental sound and impression caption pair.
Most of the captions received an appropriateness score of 3 or higher.
This result indicates that we can use ChatGPT to generate many appropriate impression captions to express each sound.

\begin{table*}[t!]
\caption{Results of audio-to-text and text-to-audio retrievals}
\label{table:result_retrieval}
\centering
\small
\begin{tabular}{@{}lrrrr|rrrr@{}}
    \wcline{1-9}
    \\[-8pt]
    \multirow{2}{*}{Method} & \multicolumn{4}{c|}{A $\rightarrow$ T} & \multicolumn{4}{c}{T $\rightarrow$ A}\\
    \cline{2-9}
     & R@1 & R@5 & R@10 & mAP@10 & R@1 & R@5 & R@10 & mAP@10\\ 
    \wcline{1-9}
    \\[-8pt]
    Random & $0.003$ & $0.014$ & $0.023$ & $0.007$ & $0.006$ & $0.011$ & $0.034$ & $0.010$ \\
    Ours & $0.034$ & $0.131$ & $0.202$ & $0.077$ & $0.031$ & $0.099$ & $0.168$ & $0.062$ \\
    \wcline{1-9}
\end{tabular}
\end{table*}
\subsection{Objective evaluation}
\label{object_eval}
\begin{figure}[t!]
\centering
\includegraphics[scale=0.93]{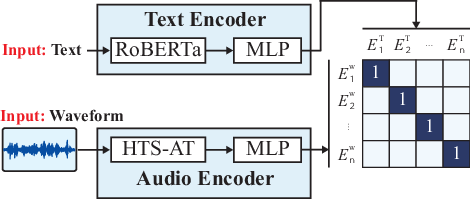}
\caption{Overview of CLAP}
\label{fig:CLAP}
\end{figure}

To evaluate our dataset objectively, we conducted text-to-audio retrieval (T $\rightarrow$ A) and audio-to-text retrieval (A $\rightarrow$ T).
First, we trained a deep learning model to obtain the correspondence between environmental sounds and impression captions using the method of contrastive language-audio pre-training (CLAP) \cite{wu_ICASSP2023_01}.
Figure~\ref{fig:CLAP} shows an overview of CLAP trained in this study. 
CLAP is trained to embed sound $E_n^W$ and text $E_n^T$ in the same vector space through contrastive learning.
We used the hierarchical token semantic audio transformer (HTS-AT) \cite{chen_ICASSP2022_01} for the audio encoder and RoBERTa \cite{liu_arXiv2019_01} for the text encoder.
For HTS-AT, we used a pre-trained model provided officially by CLAP\footnote{https://github.com/LAION-AI/CLAP}, and for RoBERTa, we used ``japanese-roberta-base''\footnote{https://huggingface.co/rinna/japanese-roberta-base}, which is trained in Japanese and provided by rinna.
When training the model, the parameters for HTS-AT and RoBERTa were fixed, and only the parameters for the multilayer perceptron (MLP) part of each audio encoder and text encoder were trained.

We used a total of 784 pairs for model training, with one caption randomly selected from the three impression captions per sound.
For the validation and test data, we used 65 and 351 environment sound--impression caption pairs, respectively, with one caption randomly selected per sound.

When performing retrieval, we used the audio and text encoders of the trained CLAP to calculate the cosine similarity between the embedding vector for the input data and the embedding set to be retrieved. 
High cosine similarity means higher retrieval results.

We used the mean average precision at the top 10 (mAP@10) and Recall at $k$ (R@$k$) as evaluation metrics. 
R@$k$ is the recall score obtained by averaging the top k retrieval results overall queries\footnote{In the case of text-to-audio retrieval, the text is the query}.

Table \ref{table:result_retrieval} shows the results.
In the table, ``random'' indicates the evaluation score on the test data before model training, and ``ours'' indicates the score on the test data after training using our impression caption dataset constructed as described in Sec.~\ref{creation_dataset}. 
Comparing the mAP@10 scores before and after model training, we confirmed that T $\rightarrow$ A and A $\rightarrow$ T performances improved by 0.070 and 0.052 points, respectively.
Similarly, we confirmed that the R@1, R@5, and R@10 scores improved after model learning.
The improvement in the scores of each evaluation metric before and after model learning indicates that the correspondence between environmental sounds and impression captions was trained to some extent and that the impression captions given were appropriate descriptions for the environmental sounds.
The relatively low scores after model learning might be due to the design of the impression captions, which did not include detailed information such as sound events, thus increasing the difficulty level compared with conventional audio--text retrieval tasks.
	    
\section{Conclusion}	
\label{conclusion}
In this study, we created a dataset with impression captions for environmental sounds that describe the impressions humans have when hearing the sounds.
We collected impression words by crowdsourcing and generated impression captions for environmental sounds using ChatGPT.
From the results of the analysis of our dataset, we confirmed that we were able to generate appropriate impression captions for environmental sounds. 
In the future, we will conduct benchmark analyses of audio captioning and text-to-audio generation using our dataset.

%

\section{Acknowledgements}
The work was supported by JSPS KAKENHI Grant Numbers 22KJ3027, 22H03639, and 23K16908, ROIS NII Open Collaborative Research 2024-(24S0504), JST Moonshot Grant Number JPMJMS2237, and JST SPRING Grants Number JPMJSP2101.
The authors also thank Maia Kuriswa for her support in collecting impression words for environmental sounds.

\ninept
\bibliographystyle{IEEEbib}
\bibliography{mybib}

\end{document}